\documentclass[preprint,superscriptaddress,nofootinbib]{revtex4-1}
\usepackage{graphicx}
\usepackage{dcolumn}
\usepackage{bm}
\usepackage{amsmath}
\usepackage{amsfonts} 
\usepackage{latexsym}
\usepackage{bbm}
\usepackage{color}
\usepackage{amssymb}
\usepackage{amsthm}
\usepackage{epsf}
\usepackage{epsfig}
\usepackage{caption3}
\usepackage{appendix}
\usepackage{cases}
\usepackage{subfigure}
\usepackage{multirow}
\makeatletter

\newcommand{\Rmnum}[1]{\expandafter\@slowromancap\romannumeral #1@}
\makeatother

\begin{document}

\title{Bubble wall velocity beyond \emph{leading-log} approximation in electroweak phase transition}
	
\author{Xiao Wang}%
\email{wangxiao2016@ihep.ac.cn}
\affiliation{Theoretical Physics Division, Institute of High Energy Physics, Chinese Academy of Sciences, 19B Yuquan Road, Shijingshan District, Beijing 100049, China}
\affiliation{School of Physics, University of Chinese Academy of Sciences, Beijing 100049, China}

\author{Fa Peng Huang}%
\email{fapeng.huang@wustl.edu}
\affiliation{Department of Physics and McDonnell Center for the Space Sciences, Washington University, St.
Louis, MO 63130, USA}
	
\author{Xinmin Zhang}
\affiliation{Theoretical Physics Division, Institute of High Energy Physics, Chinese Academy of Sciences, 19B Yuquan Road, Shijingshan District, Beijing 100049, China}
\affiliation{School of Physics, University of Chinese Academy of Sciences, Beijing 100049, China}

\bigskip
	
	
\begin{abstract}
The bubble wall velocity is essential for the phase transition dynamics in the early universe 
and its  cosmological implications, such as the energy budget of phase transition  gravitational wave and electroweak baryogenesis.
One key factor to determine the wall velocity is the collision term that quantifies the interactions between  the massive particles in the plasma and the bubble wall.
We improve the calculations of the collision term beyond the \emph{leading-log} approximation,
and further obtain more precise bubble wall velocity
for a representative effective model.
\end{abstract}
	

\maketitle
	
\section{Introduction}
The bubble wall velocity  plays  the essential roles in the energy budget~\cite{Espinosa:2010hh,Giese:2020rtr,Giese:2020znk,Wang:2020nzm,Leitao:2010yw,Leitao:2014pda} of the phase transition gravitational wave, 
electroweak baryogenesis~\cite{Cline:2020jre}, the properties of phase transition related dark matter \cite{Baker:2019ndr,Chway:2019kft,Huang:2017kzu} and so on.
Successful electroweak baryogenesis favors the deflagration hydrodynamical process where the bubble wall velocity is smaller than the sound speed.
However, the strong phase transition gravitational wave signal  prefers the supersonic bubble wall velocity which induces the detonation hydrodynamical process.
The bubble wall velocity of a given new physics model is generally determined by  the hydrodynamical and micro-physical processes which basically involve the particles scattering processes between the relevant particles of the new model at the vicinity of the bubble wall.
It is complicated to quantify  these processes precisely in the thermal plasma due to the
thermal effects and the infrared behavior. 
This makes it difficult to precisely predict the bubble wall velocity.
Most of previous studies on electroweak baryogenesis, phase transition gravitational wave, phase transition related dark matter just take the  bubble wall velocity as an input parameter. 
Therefore, there exist large theoretical uncertainties for these predictions due to the model-dependent bubble wall velocity.   
However, some pioneering works \cite{Dine:1992wr,Liu:1992tn,Ignatius:1993qn,Moore:1995si,Moore:1995ua,Moore:2000wx} show us how to calculate the wall velocity under certain assumptions.
The bubble wall velocity depends on the friction force acting on the expanding bubbles, which is further determined 
by the deviation of the massive particle populations from thermal equilibrium.
Basically, under semiclassical approximation,  the bubble wall velocity can be obtained by simultaneously solving the equation of motion for the Higgs field (or the order-parameter scalar field for the phase transition process) and the Boltzmann equations of the massive particle species. 
To solve the Boltzmann equations, it is crucial to calculate the collision terms, which quantify the particle scattering processes in the vicinity of bubble wall.
In Refs.~\cite{Moore:1995si,Moore:1995ua}, various approximations are taken to simplify the calculation of the collision terms, including the \emph{leading-log} approximation.
However, these approximations are quite rough.
Recently, there are some studies aiming to improve the calculations of the scattering process for more precise bubble wall velocity.
Ref.~\cite{Hoeche:2020rsg} resummed the logarithmic terms to all orders to include the infrared enhancement due to the emissions of soft gauge bosons in the \emph{leading-log} approximation.
And there are some new approaches to study the friction force, which can further determine the bubble wall velocity.
In this work, we improve the calculations of the collision term beyond the \emph{leading-log} approximation and include the contribution of Higgs boson to get more precise bubble wall velocity in a representative effective model.

This paper is organized as follows.
We briefly review the standard method for the calculation of the bubble wall velocity in Sec.~II. 
Then we derive the collision terms beyond \emph{leading-log} approximation in Sec.~III.
And for different species of heavy particle, the bubble wall velocity is calculated and the behavior of perturbations are also shown in Sec.~IV.
Conclusion is given in Sec.~V.

\section{Standard method}
To solve the bubble wall velocity, 
we firstly need to know the bubble dynamics which is quantified by the equation of motion (EOM) of the background field. 
Based on WKB approximation we can derive the EOM of the background  field\cite{Moore:1995ua,Moore:1995si,Espinosa:2010hh,Konstandin:2014zta}, 
\begin{equation}\label{eom0}
\Box\phi + \frac{\partial V_0(\phi)}{\partial \phi} + \sum\frac{dm^2}{d\phi}\int\frac{d^3p}{(2\pi)^32E} f(p,x) = 0 \,\,,
\end{equation}
where $\phi$ is the background field,  $V_0(\phi)$ is the zero-temperature part of effective potential $V_{\rm eff}(\phi,T)$ and the sum is over all massive particle species.
Eq.~\eqref{eom0} represents  the energy momentum conservation of the scalar-plasma system.
During a strong first-order phase transition, there exists an out-of-equilibrium regime around the expanding bubble wall.
Hence the distribution function can be expressed as a thermal equilibrium part plus a perturbation, $f \equiv f_0 + \delta f$.
Here, a default assumption is that the departure from thermal equilibrium can be regarded as a small perturbation from the thermal equilibrium.
If the deviation is large enough that is comparable with the equilibrium part, this perturbative approach might not work.
And the equilibrium distribution function for fermions and bosons are given as $f_0 = \frac{1}{\exp(E/T)\pm1}$ respectively in the \emph{plasma frame}.
The integral of the equilibrium part of the distribution functions give the thermal correction part of the effective potential $V_T(\phi,T)$.
Therefore, the EOM for the background field is
\begin{equation}\label{eomt}
\Box\phi + \frac{\partial V_{\rm eff}(\phi,T)}{\partial\phi} + \sum\frac{dm^2}{d\phi}\int\frac{d^3p}{(2\pi)^32E}\delta f(p,x) = 0\,\,,
\end{equation}
where $ V_{\rm eff}(\phi,T)$ is the thermal effective potential from a given new physics model.
The third term in Eq.~\eqref{eomt} behaves as the friction force acting on the bubble wall.
We can see that the friction force is the summation effect of the deviation of massive particle populations from thermal equilibrium.
Thus, in our following calculations, we should only consider the  contributions from massive particles. In the \emph{plasma frame}, the EOM can be further expressed as \cite{Moore:1995si,Moore:1995ua}
\begin{equation}\label{eom3}
-(1-v_w^2)\phi'' + \frac{\partial  V_{\rm eff}(\phi,T)}{\partial\phi} + \sum\frac{dm^2}{d\phi}\int\frac{d^3p}{(2\pi)^32E}\delta f(p,x) = 0 \,\,,
\end{equation}
where $v_w$ is the bubble wall velocity that we aim to calculate in this work.
And the prime mean derivatives with respect to the position coordinate.
For the stationary wall, all quantities $Q$ are functions of $x=z+v_wt$, and hence $\partial_tQ\rightarrow v_w\partial_zQ$, $\partial_zQ\rightarrow Q'$.

To proceed further, we need to know the deviation distribution part $\delta f_i$ for each massive particle population.
In principle, the  distribution deviation part for the fundamental particles are described by the quantum Liouville equation.
Under semiclassical approximation, it reduces to the Boltzmann equation.
Namely, when the WKB condition $p \gg 1/L_w$ ($p$ is the momentum of concerned particle and $L_w$ is the bubble width) is satisfied, the background field varies slowly and hence the distribution function for each particle can be approximated by the following Boltzmann equation
\begin{equation}\label{bm}
\frac{d}{dt}f = \left(\frac{\partial}{\partial t} + \dot{z}\frac{\partial}{\partial z} + \dot{p_z}\frac{\partial}{\partial p_z}\right)f = -C[f]\,\,,
\end{equation}
where $\dot{z}=p_z/E$ and $ \dot{p_z}=-(m^2)^{\prime}/(2E)$.
The distribution function $f$, which deviates from its equilibrium form, can be expressed as
\begin{equation}
f = \frac{1}{e^{(E+\delta)/T} \pm 1}\,\,,
\end{equation}
where $+$ is for fermions,  $-$ is for bosons, and 
$\delta$ is the perturbation.
Note WKB condition indicates that the Boltzmann equation only works appropriately for the particle with the momentum larger than the inverse of bubble wall width (rough estimation gives $p\gg 2$GeV in this work).
For infrared particles, it is not suitable. We also require that the scattering processes are not too frequent.
Basically, one can use the fluid approximation where several characteristic parameters are used to approximate each distribution to simplify  this integro–differential Boltzmann equation in Eq.~\eqref{bm}.
Namely, for each particle species, we parameterize the perturbation as \cite{Moore:1995si,Moore:1995ua}
\begin{equation}
\delta = -\mu - \mu_{bg}  - \frac{E}{T}(\delta T + \delta T_{bg}) - p_z(v + v_{bg})\,\,.
\end{equation}
And we take the background chemical potential perturbation as zero for simplicity.
$C[f]$ is the collision term, we will discuss it in detail in next section.
The collision term and the EOM of the background field are model-dependent.
Here we use the Higgs sextic effective model \cite{Zhang:1992fs,Grojean:2004xa} as a representative model to give a concrete model-dependent analysis.
With the leading-order thermal corrections, the effective potential for this model can be expressed as
\begin{equation}
V_{\rm eff}(\phi,T)\approx  \frac{\mu^2 + cT^2}{2}\phi^2 + \frac{\lambda}{4}\phi^4 + \frac{\kappa}{8\Lambda^2}\phi^6  \,\,,
\end{equation}
where $\Lambda/\sqrt{\kappa}$ is the effective cutoff scale and $c$ is the thermal correction which can be expressed as
\begin{equation}\label{cdim6}
c=\frac{1}{16}(g^{\prime 2}+3 g^2+4 y_t^2+4 \frac{m_h^2}{v^2}-12 \frac{\kappa v^2}{\Lambda^2}) \,\,.
\end{equation}
And $g^{\prime}$, $g$, $y_t$,  $v$ are the $U(1)$ gauge coupling, $SU(2)$ gauge coupling, top quark Yukawa coupling, and electroweak vacuum expectation value (VEV), respectively. 
Therefore, all interactions involved in this model are the same as the standard model  except for the Higgs self interactions.

In this work we consider three heavy particle species: top quark, Higgs boson, W boson (i.e. we approximate the W and Z boson as the same species after taking weighted average ).
And we treat other massive species that obtain light field-dependent mass as background\footnote{However, this approximation may not be valid, the mass of a particle in thermal plasma include the thermal effective mass and the field-dependent mass. And the thermal mass of particles with light field-dependent mass could also be large.}.
Solving the full Boltzmann equation is complicated.
Hence we truncate the full Boltzmann equation with three moments for an approximate solution.
And the three moments are chosen as $\int d^3p/(2\pi)^3$, $\int E d^3p/(2\pi)^2$, and $\int p_z d^3p/(2\pi)^3$ \cite{Moore:1995si,Moore:1995ua}.
These moments can give three coupled ordinary differential equations for the perturbations $[\mu, \delta T, v]$ of each particle species.
Using this specific truncation scheme\footnote{This truncation scheme can cause unphysical behaviors of the perturbation for the bubble wall velocity is close to the sound speed of the plasma. Ref.~\cite{Cline:2020jre,Laurent:2020gpg} choose another scheme to avoid this artifact. However, these two schemes show slight differences for subsonic wall velocity. We still use this scheme in this work.}, we can derive the fluid equation as the following compact matrix form
\begin{equation}
A\delta'+\Gamma\delta = F\,\,,
\end{equation}
where the vector of perturbations is
\begin{equation}
\mathbf{\delta}=(\mu_t,\delta T_t,Tv_t,\mu_H,\delta T_H,T v_H,\mu_W,\delta T_W,Tv_W)\,\,,
\end{equation}
and
\begin{equation}
F = \frac{v_w}{2T}\left(c_1^t(m_t^2)',c_2^t(m_t^2)',0,c_1^H(m_H^2)',c_2^H(m_H^2)',0,c_1^W(m_W^2)',c_2^W(m_W^2)',0\right)\,\,,
\end{equation}
\begin{equation}
\Gamma = \Gamma_0 + \frac{1}{\tilde{c}_4}\mathbb{M}
\end{equation}
\begin{equation}
A = \left(\begin{array}{ccc}
A_{t}&0&0\\
0&A_{H}&0\\
0&0&A_{W}\\
\end{array}\right),\quad \text{where}\quad
A_i = \left(\begin{array}{ccc}
v_wc_2^i&v_wc_3^i&\frac{1}{3}c_3^i\\
v_wc_3^i&v_wc_4^i&\frac{1}{3}c_4^i\\
\frac{1}{3}c_3^i&\frac{1}{3}c_4^i&\frac{1}{3}v_wc_4^i\\
\end{array}\right),\notag
\end{equation}
\begin{equation}
\Gamma_0 = \left(\begin{array}{ccc}
\Gamma_t&0&0\\
0&\Gamma_H&0\\
0&0&\Gamma_W\\
\end{array}\right),\quad\text{where}\quad
\Gamma_i = \left(\begin{array}{ccc}
\Gamma_{\mu1,i}&\Gamma_{T1,i}&0\\
\Gamma_{\mu2,i}&\Gamma_{T2,i}&0\\
0&0&\Gamma_{v,i}
\end{array}\right)\,\,.
\end{equation}
And $\mathbb{M}$ is
\begin{equation}
\mathbb{M} = \left(\begin{array}{ccc}
M_{tt}&M_{tH}&M_{tW}\\
M_{Ht}&M_{HH}&M_{HW}\\
M_{Wt}&M_{WH}&M_{WW}\\
\end{array}\right),\quad\text{where}\quad
M_{ij} = N_j\left(\begin{array}{ccc}
c_{3}^i\Gamma_{\mu2,j}&c_{3}^i\Gamma_{T2,j}&0\\
c_{4}^i\Gamma_{\mu2,j}&c_{4}^i\Gamma_{T2,j}&0\\
0&0&c_{4}^i\Gamma_{v,j}
\end{array}\right)\,\,.
\end{equation}
One can further assume the collision between massive particle species and light particle species in the fluid equations with opposite sign.
And the light particle species are treated as being at the same temperature and velocity.
Therefore, the background fluid equations can be expressed as:
\begin{equation}
\tilde{c}_4\left(v_w\delta T_{bg}' + \frac{v_{bg}'}{3}T\right) = N_t(\mu_t\Gamma_{\mu2,t} + \delta T_t\Gamma_{T2,t}) + \sum_{\rm bosons}N_b(\mu_b\Gamma_{\mu2,b} + \delta T_b\Gamma_{T2,b})\,\,,\notag
\end{equation}
\begin{equation}
\frac{\tilde{c}_4}{3}(\delta T_{bg}' + v_wTv_{bg}') = N_tv_tT\Gamma_{v,t} + \sum_{\rm bosons}N_bv_bT\Gamma_{v,b}\,\,,\quad\mu_{bg} = 0\,\,.
\end{equation}
Then the derivative of the background temperature and velocity can be written as:
\begin{equation}
\delta T_{bg}' = \frac{v_w(A + B) - (C+D)}{\tilde{c}_4\left(v_w^2 - \frac{1}{3}\right)}\,\,,\notag
\end{equation}
\begin{equation}
v_{bg}' = \frac{3v_w(C + D) - (A + B)}{T\tilde{c}_4\left(v_w^2 - \frac{1}{3}\right)}\,\,,
\end{equation}
where
\begin{align}
\begin{split}
&A = N_t(\mu_t\Gamma_{\mu2,t} + \delta T_t\Gamma_{T2,t})\,\,,\\
&B = \sum_{\rm bosons}N_b(\mu_b\Gamma_{\mu2,b} + \delta T_b\Gamma_{T2,b})\,\,,\\
&C = N_tv_tT\Gamma_{v,t}\,\,\\
&D = \sum_{\rm bosons}N_bv_bT\Gamma_{v,b}\,\,.
\end{split}
\end{align}
Here the heat capacity of the light degrees of freedom $\tilde{c}_4 = 78c_{4}^f + 18c_{4}^b$.
And the constants $c_i^{b/f}$ are defined as
\begin{equation}
c_i^{b/f}T^{i+1} = \int E^{i-1}(-f_0')\frac{d^3p}{(2\pi)^3}
\end{equation}
At lowest order in $m/T$, we have
\begin{equation}
c_{1}^b = \frac{\log(2T/m_b)}{2\pi^2},\quad c_2^b = \frac{1}{6},\quad c_3^b = \frac{3\zeta(3)}{\pi^2},\quad c_4^b = \frac{2\pi^2}{15}\,\,,
\end{equation}
for bosons, whereas
\begin{equation}
c_1^f = \frac{\log(2)}{2\pi^2},\quad c_2^f = \frac{1}{12},\quad c_3^f = \frac{9\zeta(3)}{4\pi^2},\quad c_4^f = \frac{7\pi^2}{60}\,\,,
\end{equation}
for fermions. Here $\zeta$ is the Riemann zeta function.

\section {Collision terms}
According to the seminal works \cite{Dine:1992wr,Liu:1992tn,Ignatius:1993qn,Moore:1995si,Moore:1995ua,Moore:2000wx} on the calculation of the bubble wall velocity, 
various improved methods \cite{John:2000zq,Megevand:2009gh,Megevand:2012rt,Megevand:2013hwa,Konstandin:2014zta,Dorsch:2018pat,Hoeche:2020rsg,Huber:2011aa,Huber:2013kj,Kozaczuk:2015owa,Friedlander:2020tnq,Mancha:2020fzw,Laurent:2020gpg,Cai:2020djd,Vanvlasselaer:2020niz,Balaji:2020yrx} to calculate the bubble wall velocity and the friction force are developed recently.
However, there are only a few of works that focus on the improvement of the calculation of the collision process during the phase transition.
These works perform the calculation of collision term  based on the \emph{leading-log} approximation.
Under this assumption, only the  $t$-channel scattering processes are considered, and only the logarithmic  terms are counted in the squared scattering amplitude in the $t$-channel scattering processes.
Besides these approximations, all the external particles are assumed to be massless and the thermal mass is only included in the denominator of the propagator in the $t$-channel process.
In this work, we concentrate on improving the calculation of these collision processes, and we recalculate the collision terms beyond these approximations.
We include the $t$-channel, $u$-channel and $s$-channel processes to derive the rigorous scattering amplitude.
All the thermal masses are taken into account.

The collision term of the Boltzmann equation can be expressed as \cite{Moore:1995si,Moore:1995ua}
\begin{equation}
C[f]=\sum_i\frac{1}{2E_p}\int\frac{d^3kd^3p'd^3k'}{(2\pi)^92E_k2E_{p'}2E_{k'}}  \bar{\Sigma}|M|^2  (2\pi)^4\delta^4(p+k-p'-k')\mathcal{P}[f_i]\,\,,\notag
\end{equation}
\begin{equation}
\mathcal{P}[f_i]=f_1f_2(1\pm f_3)(1\pm f_4) - f_3f_4(1\pm f_1)(1\pm f_2)\,\,.
\end{equation}
Hereafter, we use $\bar{\Sigma}|M|^2$ to represent the squared matrix elements. The color and spin are summed (averaged) over final states (initial states). 
$p$ is the momentum of the concerned heavy particle species with distribution function $f_1$. 
$k$ represents the momentum of the other incoming particle with distribution function $f_2$.
$p^{\prime}$ and $k^{\prime}$ denote the momenta of the outgoing particles and the corresponding distribution functions labeled as $f_3$, $f_4$ respectively.
The Mandelstam variables $s$, $t$, $u$ are defined as 
$s=(p+k)^2=(p^{\prime}+k^{\prime})^2$,
$t=(p-p^{\prime})^2=(k-k^{\prime})^2$,
and  $u=(p-k^{\prime})^2=(k-p^{\prime})^2$.
We now analyze the distribution function at first order.
We can write $f_i = 1/(\exp a_i \pm 1)$, and $1\pm f_i=f_i\exp a_i$.
Then
\begin{equation}
\mathcal{P}[f]=(e^{a_3+a_4} - e^{a_1+a_2})f_1f_2f_3f_4\,\,.
\end{equation}
In the \emph{plasma frame}, we set $a_i = (E_i-\delta_i)/T$, with small $\delta_i$; then we have
\begin{align}
\begin{split}
\exp(a_1 + a_2) &= \exp\left(\sum_iE_i/T\right)\times \exp(-\delta_1/T - \delta_2/T)\\
&\approx\exp\left(\sum_iE_i/T\right)\times(1-\delta_1/T-\delta_2/T)\,\,.
\end{split}
\end{align}
And to the first-order in $\delta$, we have
\begin{equation}
\mathcal{P}[f_i]\approx\frac{\delta_1+\delta_2-\delta_3-\delta_4}{T}f_1f_2(1\pm f_3)(1\pm f_4)\,\,.
\end{equation}

Then we need to consider the major scattering processes of the massive particles species.
In most of the previous studies, only two massive species are considered, namely, the top 
quark and W boson. Here, in our work, we include an extra massive particle species, the Higgs boson.

\begin{figure}[htbp]
\begin{center}
  \includegraphics[width=1\linewidth]{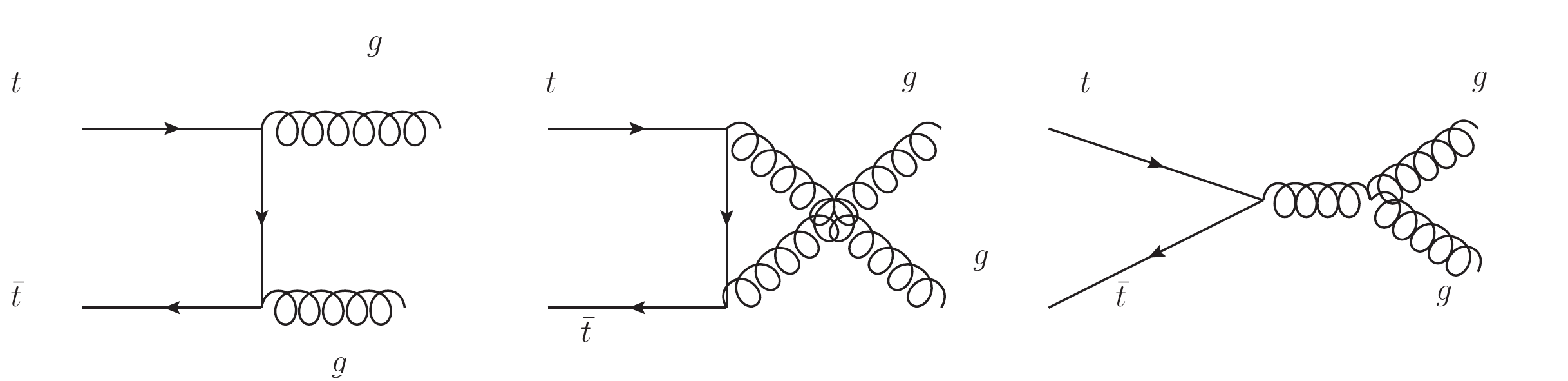}
\caption{Schematic Feynmann diagrams for top quark scattering with the bubble wall.}
\label{tg}
\end{center}
\end{figure}

For the top quark, the dominant contributions are from the strong interaction and 
we consider the interactions of  $\mathcal{O}(g_s^4)$.
The schematic Feynman diagrams contributing to the collision term of top quark are shown in Fig.~\ref{tg}.
The averaged amplitude square for the  $t\bar{t} \to g g$ at tree level can be derived as
\begin{align}
\begin{split}
\bar{\Sigma}|M|^2(t\bar{t} \to g g) &=\frac{4 g_s^4 \left(m_t^4 \left(3 t^2+14 t u+3 u^2\right)-m_t^2 \left(7 t^2 u+t^3+7 t u^2+u^3\right)-6 m_t^8+t u \left(t^2+u^2\right)\right)}{27 s^2 \left(t-m_t^2\right){}^2}\\
&\times \frac{\left(-18 m_t^2 (t+u)+18 m_t^4-s^2+9 \left(t^2+u^2\right)\right)}{\left(u-m_t^2\right){}^2}
\end{split}
\end{align}
The other channel $tg\to tg$ can be obtained by crossing symmetry.
$tq\to tq$ channel is also included in the numerical calculations.

\begin{figure}[htbp]
\begin{center}
   \includegraphics[width=0.8\linewidth]{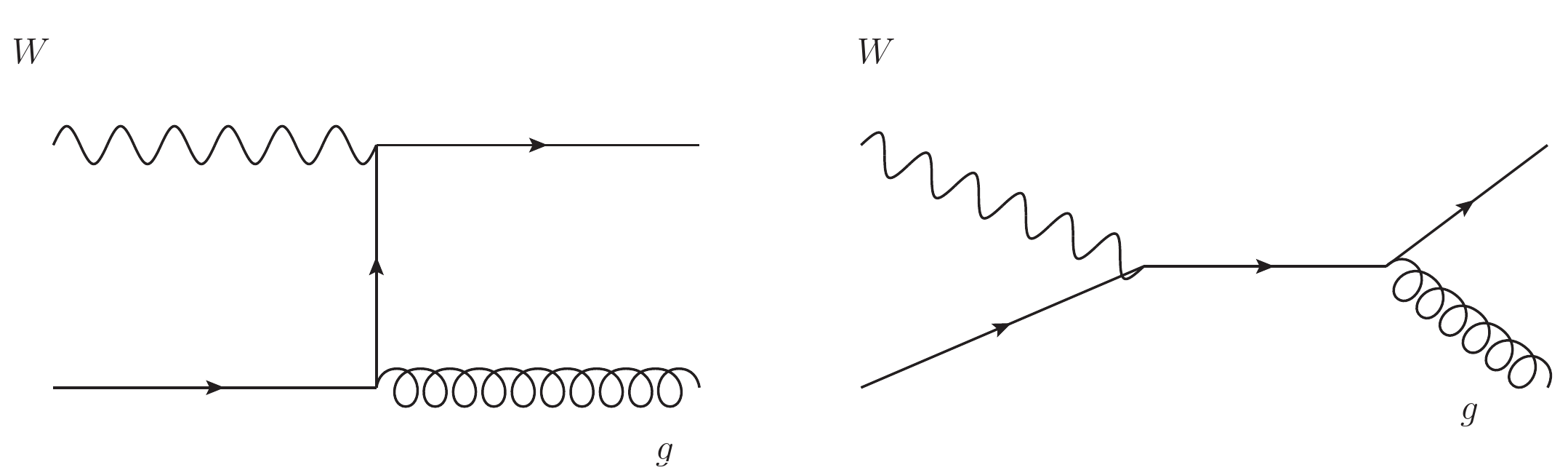}
   \caption{Schematic Feynmann diagrams for W boson scattering with the bubble wall.}\label{wq}
\end{center}
\end{figure}

For the W bosons, the dominant contributions are from the interactions of $\mathcal{O}(g_s^2 g_w^2)$.
The schematic Feynman diagrams for the collision term of W bosons are shown in Fig.~\ref{wq}.
The averaged amplitude square for the  $Wu \to d g $ at tree level can be calculated as
\begin{multline*}
\bar{\Sigma}|M|^2(W u \to d g)  =2 \text{e}^2 g_s^2 (m_d^4 (m_u^2 (4 t m_W^2+9 m_W^4+7 s^2+13 s t+10 s u+2 t^2+8 t u+u^2) \\
+t (-5 m_W^4-3 s^2-5 s t-6 s u+2 t^2-2 t u-u^2)-2 m_u^4 (4 m_W^2+5 s+5 t+2 u)+4 m_u^6)   \\
-m_d^2 (-m_u^4 (4 s m_W^2+9 m_W^4+2 s^2+13 s t+8 s u+7 t^2+10 t u+u^2)+m_u^2 (2 m_W^2 (s^2-s (6 t+u)+t (t-u)) \\
+2 m_W^4 (7 s+7 t+2 u)+s^2 (9 t+6 u)+s^3+s (9 t^2+20 t u+2 u^2)+t (t^2+6 t u+2 u^2))\\
+t (-2 m_W^2 (s^2-s (4 t+u)+t (t-u))+m_W^4 (-(6 s+5 t+4 u))-s^2 (5 t+4 u)+s^3-s (t^2+8 t u+2 u^2)\\
+t^3-t u^2)+m_u^6 (s+5 t+4 u))+m_d^6 (-m_u^2 (5 s+t+4 u)+4 m_u^4+t (3 s-t+2 u))\\
+s (m_u^4 (-5 m_W^4+2 s^2-5 s t-2 s u-3 t^2-6 t u-u^2)+m_u^2 (2 m_W^2 (s^2-s (4 t+u)+t (t-u))\\
+m_W^4 (5 s+6 t+4 u)+s^2 t-s^3+s (5 t^2+8 t u+u^2)+t (-t^2+4 t u+2 u^2))+t (-2 m_W^2 (s^2-s (2 t+u)+t (t-u))\\
+m_W^4 (-(s+t+4 u))-s^2 t+s^3-s (t^2+4 t u+u^2)+t^3-t u^2)\\
+m_u^6 (-s+3 t+2 u)))/(3 m_W^2(s-m_d^2){}^2(t-m_u^2){}^2(\sin (\theta _W)){}^2)
\end{multline*}
The $W g \to \bar{u} d$ can also been obtained by the crossing symmetry from the above expression.
We also include the interaction of $\mathcal{O}(g_w^4)$ in our numerical calculations, such as $W W\to q\bar{q}$ and $W q\to W q$.

\begin{figure}[htbp]
\begin{center}
\includegraphics[width=1\linewidth]{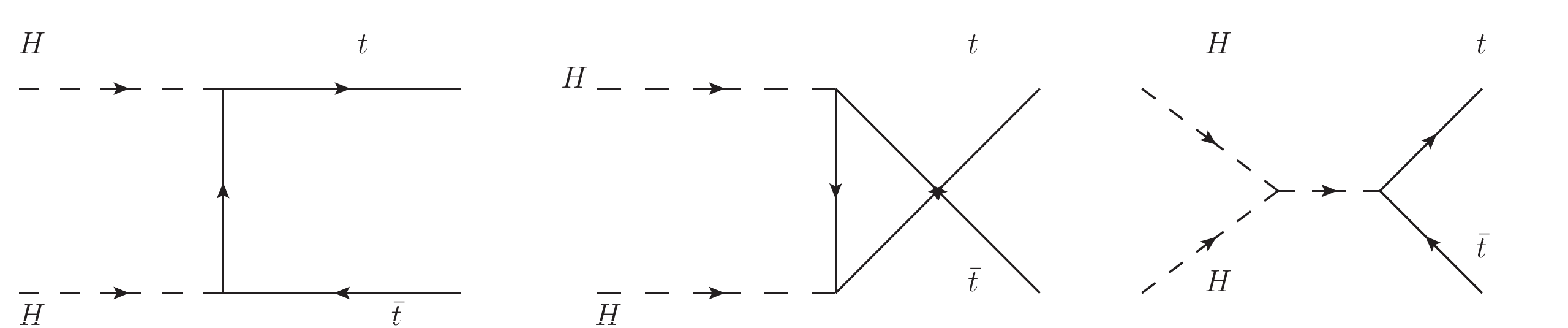}
\caption{Schematic Feynmann diagrams for Higgs boson scattering with the bubble wall.}
\label{ht}
\end{center}
\end{figure}
For the Higgs boson, the schematic Feynman diagrams for the Higgs species are shown in Fig.~\ref{ht}.
And the dominant contributions are from the interaction of $\mathcal{O}(y_t^4)$.
The averaged amplitude square for the  $HH \to t\bar{t} $ at tree level can be calculated as
\begin{multline*}
\bar{\Sigma}|M|^2(HH \to t\bar{t} ) =\text{e}^4 m_t^2      (m_H^4      (m_t^2      (3 s^2 (t-u)^2+2 s (t+u)      (t^2-(12 \delta_H +17) t u+u^2     ) \\
-t u      ((3 \delta_H +5) t^2+2 (15 \delta_H +13) t u+(3 \delta_H +5) u^2     )     )-m_t^6      (2 (18 \delta_H +35) s (t+u)\\
+(63 \delta_H +113) t^2+18 (13 \delta_H +19) t u+(63 \delta_H +113) u^2     )+m_t^4      (s      (3 (5 \delta_H +7) t^2 \\
+20 (3 \delta_H +5) t u+3 (5 \delta_H +7) u^2     )+3 (t+u)      ((\delta_H +3) t^2+2 (15 \delta_H +17) t u+(\delta_H +3) u^2     )     )\\
+m_t^8 (7 (3 \delta_H +7) s+8 (18 \delta_H +35) (t+u))-28 (3 \delta_H +7) m_t^{10}+9 (\delta_H +1) s t^2 u^2     )\\
-s m_H^2 m_t^2      (-m_t^4      (2 (9 \delta_H +34) s (t+u)+(27 \delta_H +100) t^2+18 (5 \delta_H +12) t u\\
+(27 \delta_H +100) u^2     )+m_t^2      (2 s      (3 (\delta_H +3) t^2+2 (6 \delta_H +17) t u+3 (\delta_H +3) u^2     )\\
+3 (t+u)      ((\delta_H +4) t^2+2 (3 \delta_H +4) t u+(\delta_H +4) u^2     )     )+4 m_t^6 ((3 \delta_H +14) s+2 (9 \delta_H +34) (t+u))\\
-16 (3 \delta_H +14) m_t^8+s^2 (t-u)^2+s (t+u)      (t^2-2 (3 \delta_H +7) t u+u^2     )-(3 \delta_H +4) t u (t-u)^2     )\\
-m_H^6 m_t^2 (t-u)^2      (-2 m_t^2+3 s+t+u     )+m_H^8 m_t^2 (t-u)^2+s^2 m_t^2      (-m_t^4      (16 s (t+u)+23 t^2+18 t u+23 u^2     )\\
+16 m_t^6 (s+4 (t+u))+m_t^2 (t+u)      (4 s (t+u)+3 (t-u)^2     )\\
-64 m_t^8+t u (t-u)^2     )     )/     (288          (t-m_t^2     ){}^2     m_W^4          (s-m_H^2     ){}^2          (u-m_t^2     ){}^2          (     \sin (\theta _W     )     ){}^4     )
\end{multline*}
where $\delta_H\sim (0.6 ,1.5)$ is the deviation of the triple Higgs coupling in the Higgs sextic model.
$\delta_H=0.76$ for  $\Lambda/\sqrt{\kappa}=780$~GeV.

In the electroweak plasma,
the thermal mass should also be taken into account.
For quarks, the thermal mass is $m_{qT}^2=g_s^2 T^2/6$.
For gluons, the thermal mass is $m_{gT}^2=2g_s^2 T^2$.
For W bosons, the thermal mass is $m_{WT}^2= 11g_w^2 T^2/6$.
Based on the squared matrix elements obtained at zero temperature, we add the thermal mass to  the 
zero-temperature mass as the total mass in the thermal bath. After replacing with the 
more appropriate mass term, we numerically calculate the collision term using the exact squared matrix elements and
Monte-Carlo approach instead of the analytical method under \emph{leading-log} approximation where a lot of terms are omitted. 
The Monte-Carlo package VEGAS is used in our integration calculations.
We take an infrared cutoff $p=2$~GeV to avoid the the invalidity of the Boltzmann equation.
And we can basically derive the following results
\begin{align}
\begin{split}
&\int \frac{d^3 p}{(2\pi)^3 T^2} C[f]=\mu  \Gamma_{\mu_1}+\delta T~ \Gamma_{T1}\\
&\int \frac{d^3 p}{(2\pi)^3 T^3} E C[f]=\mu  \Gamma_{\mu_2}+\delta T~ \Gamma_{T2}\\
&\int \frac{d^3 p}{(2\pi)^3 T^3} p_z C[f]=v T  ~\Gamma_{v}\,\,,
\end{split}
\end{align}
for each particle species.
Based on  above  squared matrix elements and definitions of the $\Gamma$ matrix elements, we can obtain the following $\Gamma$ matrix element at the nucleation temperature $T_n \approx 100~$GeV for $\Lambda/\sqrt{\kappa}=780$~GeV.
For the top quark, we have 
\begin{align}
\begin{split}
&\Gamma_{\mu_1,t}=4.5 \times 10^{-3} T,\quad \Gamma_{T_1,t}=3.0  \times 10^{-2} T, \\
&\Gamma_{\mu_2,t}=8.5 \times 10^{-2} T,\quad \Gamma_{T_2,t}=1.1  \times 10^{-1} T,\\
&\Gamma_{v,t}=3.6  \times 10^{-2} T.
\end{split}
\end{align}
And for the W/Z gauge bosons:
\begin{align}
\begin{split}
&\Gamma_{\mu_1,W}=2.3 \times 10^{-3} T,\quad \Gamma_{T_1,W}=4.1 \times 10^{-2} T, \\
&\Gamma_{\mu_2,W}=6.9 \times 10^{-2} T,\quad \Gamma_{T_2,W}=3.3  \times 10^{-2} T,\\
&\Gamma_{v,W}=1.7 \times 10^{-2} T.\\
\end{split}
\end{align}
For the Higgs boson, we have
\begin{align}
\begin{split}
&\Gamma_{\mu_1,H}=0.2 \times 10^{-3} T,\quad \Gamma_{T_1,H}=0.5  \times 10^{-3} T,\\
&\Gamma_{\mu_2,H}=0.5 \times 10^{-3} T,\quad \Gamma_{T_2,H}=0.9  \times 10^{-2} T,\\
&\Gamma_{v,H}=0.9  \times 10^{-3} T.\\
\end{split}
\end{align}

\section{Bubble wall velocity}
Substitute the results obtained from Boltzmann equation under semiclassical and fluid approximation into Eq.~\eqref{eom3}.
At leading order of perturbations, the EOM of Higgs can be approximated as
\begin{align}
\begin{split}
-(1-v_w^2)\phi''&+ \frac{\partial V_{\rm eff}(\phi,T)}{\partial\phi} + \delta T_{\rm bg}\frac{\partial^2 V_{\rm eff}(\phi,T)}{\partial T\partial\phi} \\
&+ \frac{N_tT}{2}\frac{dm_t^2}{d\phi}\left(c_{1}^f\mu_t + c_{2}^f(\delta T_t + \delta T_{bg})\right)\\
&+ \frac{N_WT}{2}\frac{dm_W^2}{d\phi}\left(c_{1}^b\mu_W + c_{2}^b(\delta T_W + \delta T_{bg})\right)\\
&+ \frac{N_HT}{2}\frac{dm_H^2}{d\phi}\left(c_{1}^b\mu_H + c_{2}^b(\delta T_H + \delta T_{bg})\right) = 0 \label{eom}
\end{split}
\end{align}
where $N_i$ the degree of freedom of particle $i$, and the temperature $T$ should actually be $z$-dependent, namely, $T = T_+ + \delta T_{bg}(z)$. Here $T_+$ is temperature just in front of the bubble wall and can be solved with the hydrodynamical treatment\footnote{Previous studies \cite{Wang:2020nzm,Huber:2013kj,Friedlander:2020tnq} show the temperature profile can be derived from a specific equation of state. However, the most appropriate approach to derive the temperature profile of a concrete particle physics model should begin with the effective potential as mentioned in Ref.~\cite{Wang:2020nzm}.} of expanding bubble \cite{Wang:2020nzm,Espinosa:2010hh,Leitao:2010yw,Leitao:2014pda,Giese:2020rtr,Giese:2020znk}.
In this work, we take the approximation $T_n = T_+ = T$ in Eq.~\eqref{eom} for simplicity.
To proceed further, the wall shape ansatz is chosen as
\begin{equation}
\phi(z) = \frac{\phi_0}{2}\left(1 + \tanh\frac{z + v_wt}{L_w}\right)\,\,,\label{bubble}
\end{equation}
where $\phi_0$ and $L_w$ are the VEV of the Higgs boson in the broken phase and the bubble wall width respectively.
And this ansatz may only appropriate for the weak first-order phase transition, since the bubble profile is quite similar to Eq.~\eqref{bubble} in that situation.
However, this ansatz might not be suitable for the bubble profile of very strong first-order phase transition, such as the strong supercooling and the ultra supercooling cases mentioned in Ref.~\cite{Wang:2020jrd}.
In this work, we consider a benchmark point with relatively weak first-order phase transition and we assume this shape in the following analysis.

For a given specific bubble wall shape in Eq.~\eqref{bubble}, it is still difficult to fully solve the EOM in Eq.~\eqref{eom}.
Hence, for an approximate solution, the following two constraints \cite{Moore:1995si,Moore:1995ua} should be satisfied
\begin{equation}
\int [\text{Eq.\eqref{eom}}]\phi'dz=0,\quad\int [\text{Eq.\eqref{eom}}]\frac{z}{L_w}\phi'dz=0\,\,.\label{moments}
\end{equation}
The first equation in Eq.~\eqref{moments} means the total pressure on the bubble wall vanishes in the steady velocity regime.
And the second equation indicates the bubble wall width should not change, that means the asymmetry in the total pressure between the front and behind the bubble wall should be zero for the steady bubble wall velocity.

\begin{table}[t]
	\centering
	\begin{tabular}{cc|cccc}
		\hline\hline
		&$\Lambda/\sqrt{\kappa}$~[GeV]&$ v_w$&$L_wT$&$v_c/T_c$&$T_n$~[GeV]\\
		\hline
		Two-particle &780&0.3382&20.1863&1.1044&100.977\\
		Three-particle &780&0.2499&18.1759&1.1044&100.977\\
		\hline\hline
	\end{tabular}
	\caption{Bubble velocity and phase transition parameters for benchmark point with different kinds of particle species counted in the EOM.}\label{vtable}
\end{table}
From the above set-up and the collision term derived in the last section, we calculate the bubble velocity for different species of heavy particle that are counted in the Higgs sextic effective model.
In Tab.~\ref{vtable}, we show the bubble wall velocity, bubble width, wash-out parameter, and nucleation temperature for a benchmark point with $\Lambda/\sqrt{\kappa} = 780\rm~ GeV$.
For the two-particle case which only considers top quark and W boson, we use the collision term derived in Ref.~\cite{Moore:1995ua} to obtain the bubble wall velocity.
And the bubble wall velocity for three-particles case is derived with our calculation of the collision term beyond the \emph{leading-log} approximation.
From Tab.~\ref{vtable}, we can see the bubble wall velocity derived with our results of collision term is small than the velocity derived with the results of Ref.~\cite{Moore:1995ua}.
The lower bubble wall velocity originates from more rigorous scattering processes, which effectively increase the friction force similar to the behavior shown in Ref.~\cite{Hoeche:2020rsg}.

\begin{figure}[htbp]
	\centering
	\subfigure{
		\begin{minipage}[t]{0.5\linewidth}
			\centering
			\includegraphics[scale=0.55]{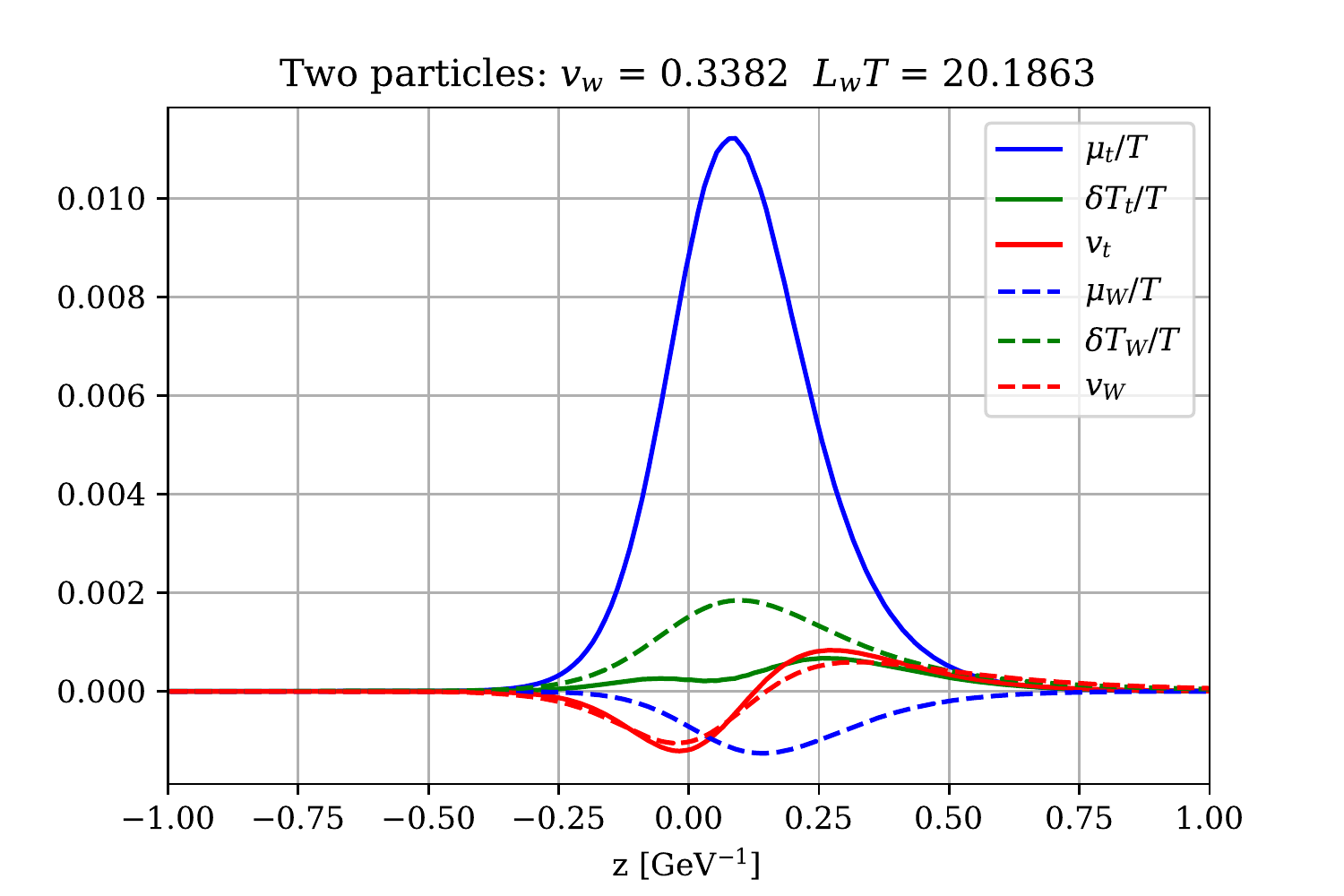}
	\end{minipage}}%
	\subfigure{
		\begin{minipage}[t]{0.5\linewidth}
			\centering
			\includegraphics[scale=0.55]{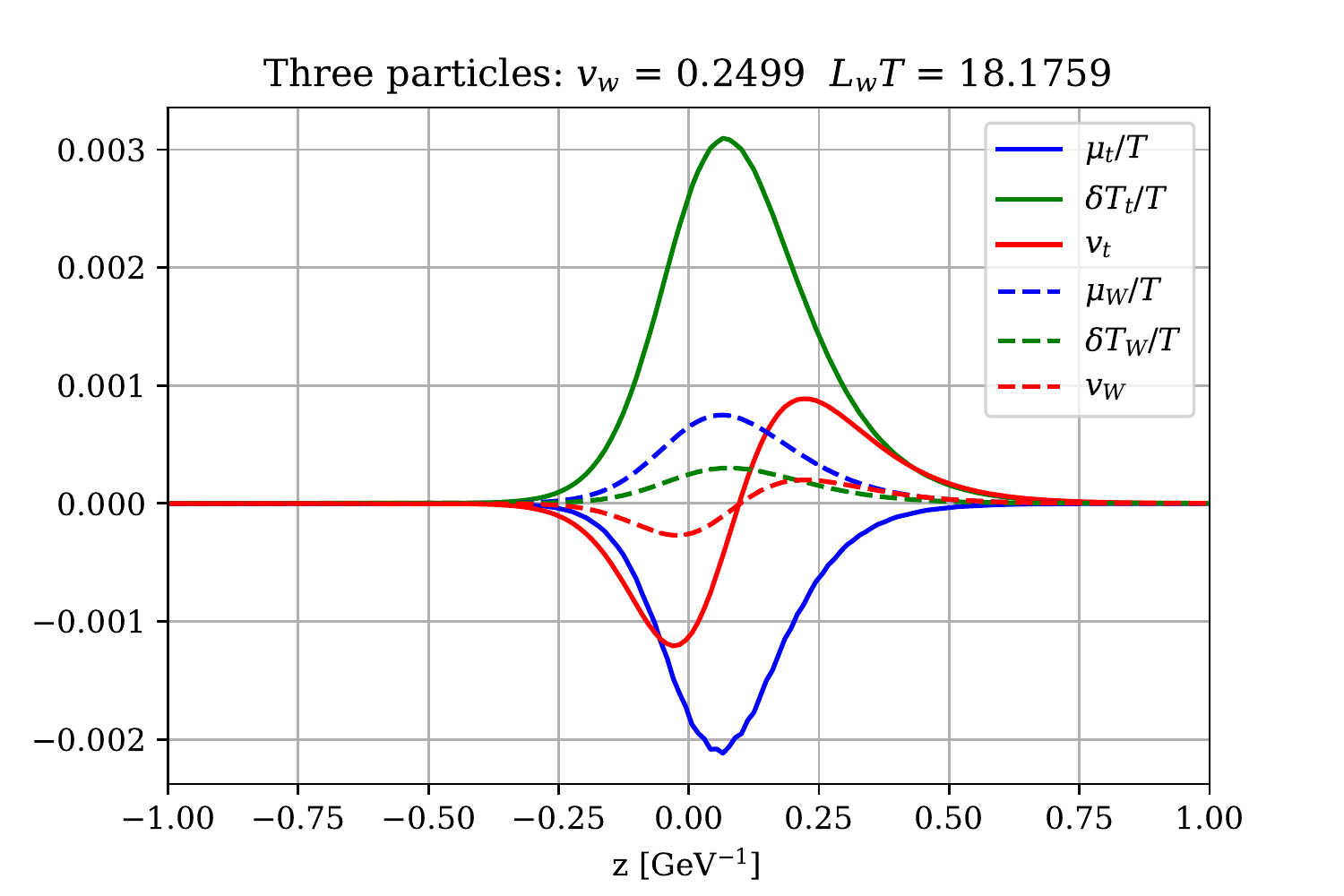}
	\end{minipage}}	
	\caption{An illustration of perturbations for top qurk and W boson. Left: the perturbations of top quark (solid line) and W boson (dashed line) for two-particle species. Right: the perturbations of top quark (solid line) and W bosons (dashed line) for three-particle species.}\label{perturb1}
\end{figure}

\begin{figure}[htbp]
	\centering
	\subfigure{
		\begin{minipage}[t]{0.5\linewidth}
			\centering
			\includegraphics[scale=0.55]{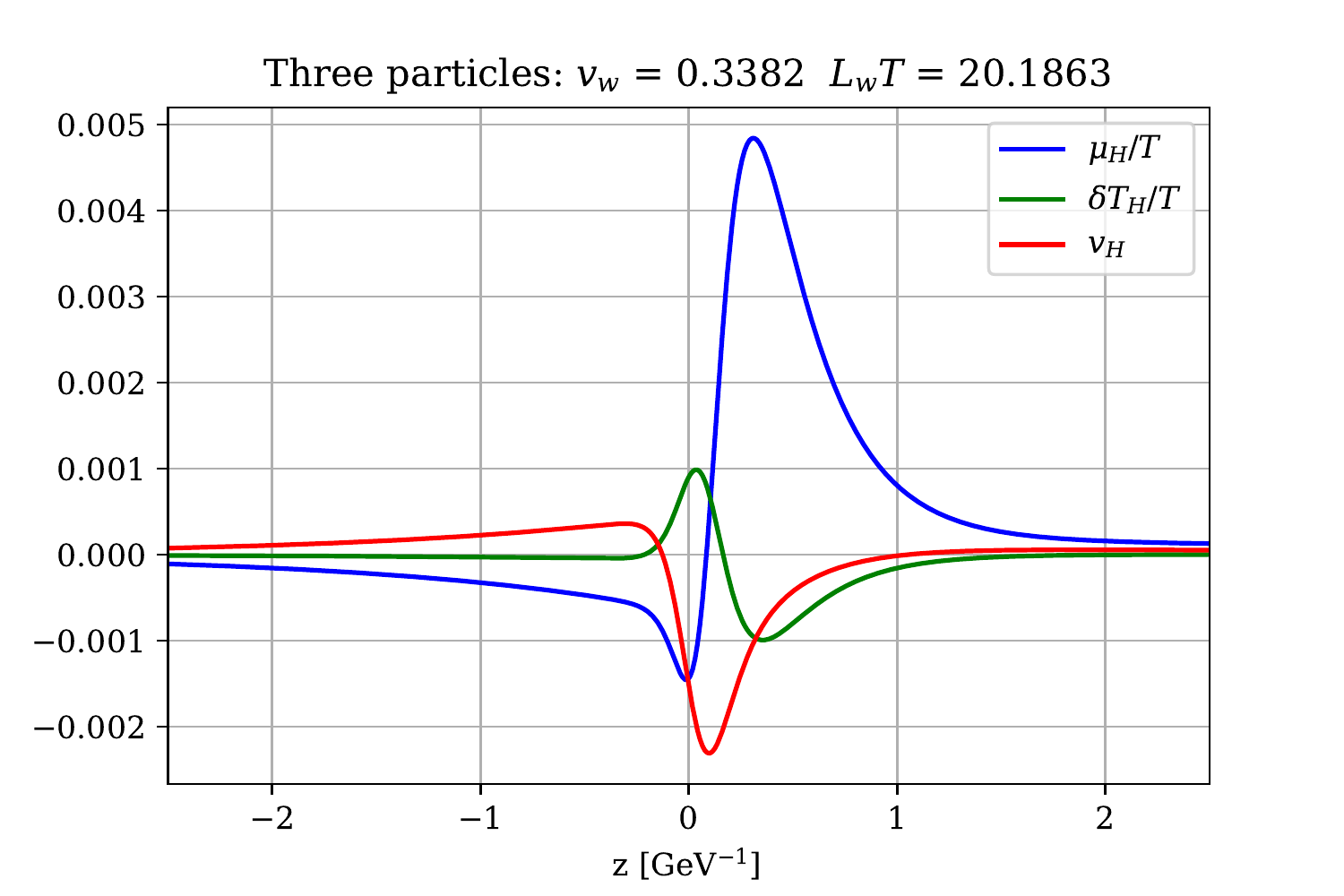}
	\end{minipage}}%
	\subfigure{
		\begin{minipage}[t]{0.5\linewidth}
			\centering
			\includegraphics[scale=0.55]{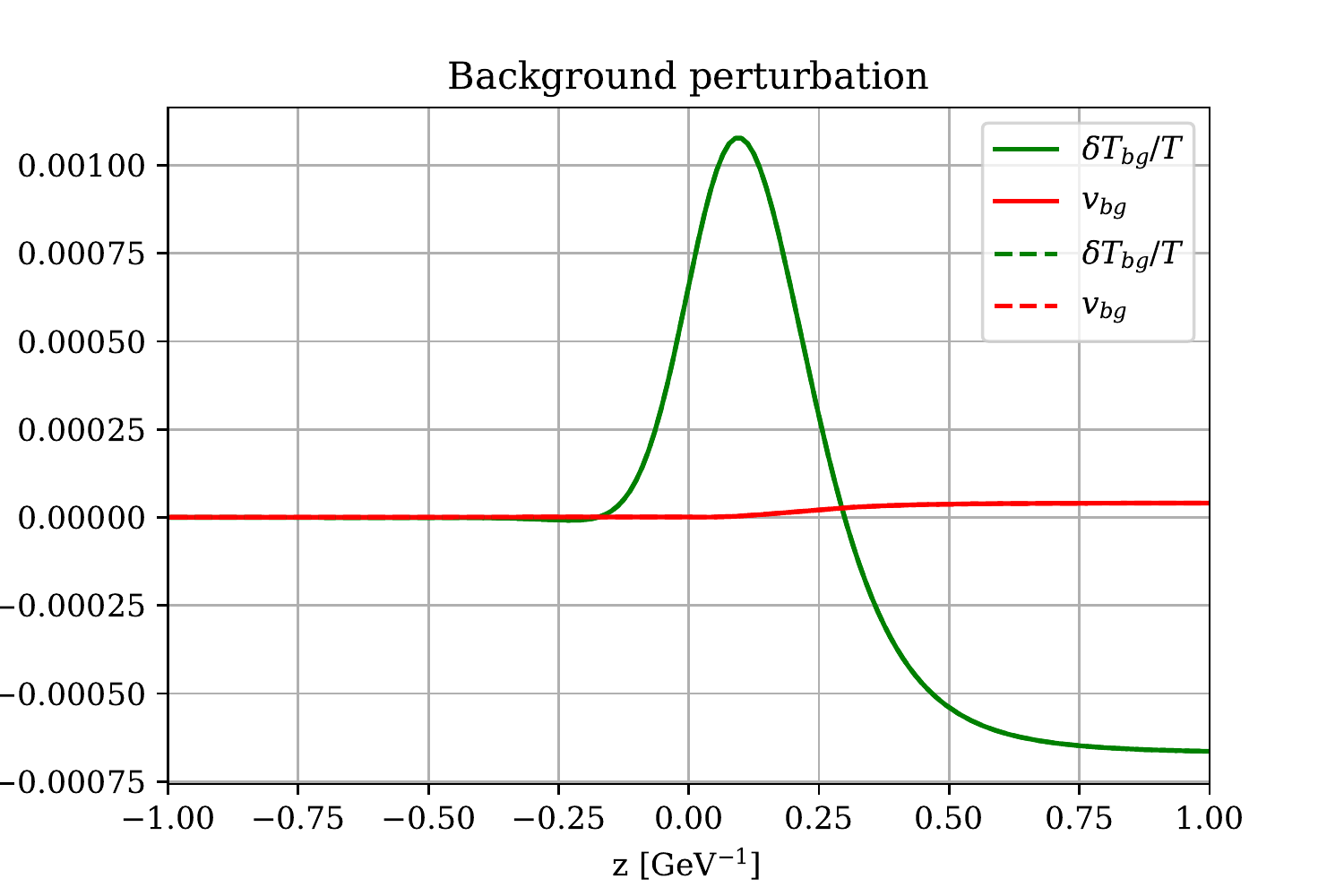}
	\end{minipage}}	
	\caption{An illustration of perturbations for Higgs boson and background particle species. Left: the perturbation of Higgs boson for three-particle species. Right: the background perturbations for three-particle species (solid line) case and two-particle species (dashed line) case.}\label{perturb2}
\end{figure}

For the bubble velocity of both cases, we show the perturbations of different particles in Fig.~\ref{perturb1} and Fig.~\ref{perturb2}.
The chemical potential perturbation, the temperature perturbation, and the velocity perturbation are denoted by blue, green, and red lines, respectively.
The left panel of Fig.~\ref{perturb1} shows the temperature normalized perturbations of top quark and W boson with solid and dashed line respectively, for two-particle species.
And the perturbations of top quark and W bosons for three-particle species are depicted in the right panel of Fig.~\ref{perturb1}.
According to this figure, we find the chemical potential perturbation and temperature perturbation of top quark and W boson are significantly changed. 
The chemical potential perturbations of three-particle case have opposite sign compared with the two-particle case for top quark and W boson.
Especially, for the top quark, its chemical potential perturbation of three-particle case is about one order of magnitude smaller than its values of two-particle case.
However, the values of chemical potential perturbation of both cases for W boson are basically at the same order of magnitude.
In the three-particle case, the temperature perturbation of top quark is amplified.
Instead for the W boson, the temperature perturbation is reduced.
The velocity perturbations of top quark and W boson are barely affected by different heavy species that are counted in calculation.
In Fig.~\ref{perturb2}, we show the perturbations of Higgs boson in the left panel for three-particle case, and the result of background particle species are plotted in the right panel.
Here we use the solid and dashed line to represent the background perturbation for two-particle and three-particle case respectively.

\section{Discussion and Conclusion}
The precise prediction of bubble wall velocity is crucial for our understanding of 
the phase transition dynamics and its cosmological implications.
For example, the value of bubble wall velocity is strongly related to the hydrodynamical mode around the expanding bubble wall; 
for bubble wall velocity lower than the sound speed, it is the deflagration mode, which 
is favored by the electroweak baryogenesis;
for the wall velocity much larger than the sound speed (basically, it should be larger than the Jouguet detonation velocity), it is called the detonation mode,
which could produce much stronger phase transition gravitational wave compared to the
deflagration case.
Therefore, it is important to precisely calculate the bubble wall velocity.
The main difficulty is the precise calculations of the collision terms, which represent
the various particle scattering processes at the vicinity of the bubble wall.
In previous studies, only the $t$-channel process and the logarithmic term of the $t$-channel are considered.
All particle masses are assumed massless except for the propagator of the $t$-channel process.

In this work, we have considered the complete scattering amplitude of all the channels  instead of only logarithmic terms of the $t$-channel.
And all the thermal masses are included in our numerical calculations.
The contribution of massive Higgs boson is also taken into account.
Using Monte-Carlo algorithm, we have numerically calculated more precise collision term and
hence obtained more reliable bubble wall velocity in the representative effective model.
After considering more rigorous scattering processes and the corresponding collision terms,
our  bubble wall velocity is smaller than the previous result under the \emph{leading-log} approximation.
It could help us to improve the prediction on the baryogenesis and the gravitational wave spectra.
Based on the recent studies~\cite{Arnold:2000dr,Arnold:2002zm,Arnold:2003zc,Bodeker:2017cim}, more precise calculations 
including the resummation over the large logarithmic terms~\cite{Hoeche:2020rsg}
are left for our future study.

\begin{acknowledgments}
FPH is supported in part by the McDonnell Center for the Space Sciences. 
XW and XMZ are supported in part by  the Ministry of Science and Technology of China (2016YFE0104700), the National Natural Science Foundation of China (Grant NO. 11653001), the CAS pilot B project (XDB23020000).
\end{acknowledgments}


\end{document}